\documentclass[letterpaper,aps,prd,preprint,showpacs,nofootinbib,superscriptaddress]{revtex4-1}
\usepackage{dcolumn}
\usepackage{placeins}
\usepackage{graphicx}
\usepackage{color}
\usepackage{amsmath}
\usepackage{amssymb}
\usepackage{amsthm}
\usepackage{latexsym}
\usepackage{bm}
\usepackage{multirow}
\usepackage{cancel}
\usepackage{float}
\usepackage{hyperref}
\usepackage{dsfont} 

\definecolor{tableHeader}{gray}{0.5}
\definecolor{tableShade}{gray}{0.9}
\usepackage{tabularx}
\usepackage[table]{xcolor}
\newcolumntype{R}{>{\raggedright\arraybackslash}X}
\newcolumntype{C}{>{\centering\arraybackslash}X}

\usepackage[
labelfont=sf,
hypcap=false,
format=hang,
width=0.8\columnwidth]{caption}

\providecommand{\openone}{\leavevmode\hbox{\small1\kern-3.8pt\normalsize1}}

\newcommand{\bea}{\begin{eqnarray}}
\newcommand{\eea}{\end{eqnarray}}
\def\AZH{\(A \rightarrow ZH\)}
\def\Hbb{\(H \rightarrow b\bar{b}\)}
\def\Abb{\(A \rightarrow b\bar{b}\)}
\def\HZA{\(H \rightarrow ZA\)}

\begin{document}
\vspace*{3.0truecm}

\title{Mapping  $pp\to A\to ZH\to l^+l^-b\bar b$ and $pp\to H\to ZA\to l^+l^-b\bar b$ \\ Current and Future Searches  onto 2HDM Parameter Spaces}

\author{Rachid Benbrik}
\email[]{r.benbrik@uca.ac.ma}
\affiliation{\small Laboratoire de Physique fondamentale et Appliq\'ee Safi, Facult\'e Polydisciplinaire de Safi, Sidi Bouzid, BP 4162,  Safi, Morocco}

\author{Henry Day-Hall}
\email[]{h.day-hall@soton.ac.uk}
\affiliation{\small School of Physics and Astronomy, University of Southampton, Southampton, SO17 1BJ, United Kingdom}
\affiliation{\small Particle Physics Department, Rutherford Appleton Laboratory, Chilton, Didcot, Oxon OX11 0QX, United Kingdom}
\author{Stefano Moretti}
\email[]{s.moretti@soton.ac.uk}
\affiliation{\small School of Physics and Astronomy, University of Southampton, Southampton, SO17 1BJ, United Kingdom}
\author{Souad Semlali}
\email[]{s.seemlali@gmail.com}
\affiliation{\small Laboratoire de Physique fondamentale et Appliq\'ee Safi, Facult\'e Polydisciplinaire de Safi, Sidi Bouzid, BP 4162,  Safi, Morocco}

\date{\today}

\vspace*{-3cm}

\begin{abstract}
    By borrowing the results from a Large Hadron Collider (LHC) analysis performed with \(36.1~\text{fb}^{-1}\) of Run 2 data 
    intended to search for \(A\) production followed by \(ZH\) decay in turn  yielding \(l^+l^-b\bar b\) (\(l=e,\mu\)) final states in the context of 
 the standard four Yukawa types of the 2-Higgs Doublet Model (2HDM), we recast it in terms of sensitivity reaches for the similar process \(pp\to H\to ZA\to l^+l^-b\bar b\). This simple exercise across the two processes, which is possible because the only kinematic difference between these are different widths for the Higgs bosons, in turn affecting minimally the efficiency of an experimental selection, enables us to  expand the region of parameter space that can be tested to the case when $m_H\ge m_A+m_Z$. Furthermore, we extrapolate our results to full Run 3 data samples. We conclude that, while the high energy and luminosity stage of the LHC can afford one with increased sensitivity to the 2HDM in general, the recast analysis  does not add anything to what already probed through the actual one.
\end{abstract}

\maketitle

\section{Introduction}

Following the discovery of a 125 GeV Higgs boson at the Large Hadron Collider (LHC), several studies of its properties have been carried out over the years.
The  situation at present is that the measured Higgs signal rates in all accessed production and decay channels agree with the Standard Model (SM) predictions.
Although the current LHC Higgs data are generally consistent with the SM,
the possibility that the observed Higgs state could be part of a model with an extended Higgs dynamics,
one that includes an extra doublet, still exists.
Therefore, since the discovered Higgs state belongs to a doublet,
one is induced to consider a generic 2-Higgs Doublet Model (2HDM)~\cite{Branco:2011iw}.

This Beyond the SM (BSM) scenario contains two Higgs doublets used to
give mass to all gauge bosons and fermions of the SM.
The Higgs particle spectrum of the 2HDM is as follows:
two CP even (\(h\) and \(H\), with, conventionally, \(m_h < m_H\)),
one CP odd (\(A\))
and a pair of charged (\(H^\pm\)) Higgs bosons.
Amongst the many signals that these additional Higgs states could produce,
of particular relevance are those involving their cascade decays,
wherein a heavier Higgs state decays in a pair of lighter ones or else into a light Higgs state and a gauge boson.
This is the case as the former process gives access to the shape of the Higgs potential of the enlarged Higgs sector
while the latter channel is intimately related to the underlying gauge structure, which may well be larger than the SM one. 

We concern ourselves here with the second kind of processes,
specifically involving only the neutral Higgs states in addition to the discovered SM-like one, {which} in our 2HDM is identified with the \(h\) state.
In short, we intend to study \(A\to ZH\) and \(H\to ZA\) decays\footnote{The case of the corresponding charged Higgs boson decays of the type \(H^\pm\to W^\pm H\) and \(W^\pm A\) has been recently reviewed in~\cite{Akeroyd:2016ymd}.}.
The pattern of Branching Ratios (BRs) of the two decays \(A\to ZH\) and \(H\to ZA\) was first discussed in Refs.~\cite{Moretti:1994ds} and~\cite{Djouadi:1995gv}
(albeit in a Supersymmetric version of the 2HDM) and more recently implemented in Refs.~\cite{Djouadi:1997yw,Krause:2018wmo} in the 2HDM.
As for production channels, the by far most relevant one  is gluon-gluon fusion, i.e., \(gg\to A\) or \(H\),
with an {occasional} competing contribution from \(b\bar b\to A\) or \(H\), respectively. 

LHC searches for the complete channels \(gg,b\bar b\to A\to ZH\) and \(gg, b\bar b\to H\to ZA\) have been carried out at both ATLAS~\cite{Aaboud:2018eoy} and CMS ~\cite{Khachatryan:2016are,Sirunyan:2019wrn},
by exploiting leptonic decays of the gauge boson, \(Z\to l^+l^-\) (\(l=e,\mu\)),  and hadronic decays of
the accompanying neutral Higgs state, in particular, \(H\) or \(A\to b\bar b\) or \(\tau^+\tau^-\).
Based on this approach, 
current experimental data exclude heavy neutral Higgses with masses up to about \(600\)--\(700\) GeV,
depending on the BSM Higgs spectrum and the value of \(\tan(\beta)\), 
the ratio of the Vacuum Expectation Values (VEVs) of the aforementioned two Higgs doublets.
These findings are broadly in line with previous phenomenological results obtained in 
Ref.~\cite{Coleppa:2014hxa}, which had forecast the LHC scope in accessing both \(A\to ZH\) and \(H\to ZA\) decays in a variety of final states. 

Far away from the alignment limit, $\sin(\beta-\alpha) = 1$, searches have been carried out at the LHC Run 2 looking for additional Higgs bosons decaying to $A\to hZ$ or/and $H\to hh$ leading to $l^+l^- b\bar b$ \cite{Sirunyan:2019xls, Aaboud:2017cxo} or/and $\tau^+\tau^- b\bar b$ \cite{Sirunyan:2019xjg}. While in the exact alignment limit, $A\to hZ$ and $H\to hh$ will be suppressed, $A/H \to H/A Z$ is unsuppressed if kinematically open. There are additional reasons for studying \(A\to ZH\) and \(H\to ZA\) decays.
For a start, Ref.~\cite{Ferreira:2017bnx} emphasised the importance of using the \(pp\to A\to Zh\) process to test the wrong-sign limit of the so-called 2HDM Type-II (see below). 
Furthermore, Ref.~\cite{Dorsch:2014qja} highlighted the fact that this very same process echoes the dynamics of the EW Phase Transition (EWPT).
It is the scope of this paper to revisit these two decay chains, in particular,
we intend to use a synergetic approach that recasts the results of experimental searches in one mode,
interpreted in terms of 2HDM constraints, into the scope of the other in the same respect.
This is possible because they can have the same final state.
Here, we consider the final state \( l^+l^- b\bar b\) and start from the results of~\cite{Aaboud:2018eoy}
for the \(A\to ZH\) decay in order to obtain the corresponding ones for the complementary channel \(H\to ZA\),
altogether showing that such a recasting can afford one with a much stronger sensitivity that either channel alone can offer.

The plan of the paper is as follows.
In the next section, we introduce the 2HDM.
We then scan its parameter space in order to establish the sensitivity of LHC data analyses to such a BSM scenario and map the findings of one channel into the other.
We then conclude.
\section{The 2HDM with theoretical and experimental constraints}
\subsection{The Model}
Unlike the SM, the 2HDM contains two complex scalar doublets \(\Phi_{1,2}\) from \(SU(2)_L\) with 
the most general gauge invariant renormalisable scalar potential of the 2HDM  given by:
\begin{align}\nonumber
V(\Phi_1,\Phi_2) =& m_{11}^2\Phi_1^\dagger\Phi_1+m_{22}^2\Phi_2^\dagger\Phi_2-[m_{12}^2\Phi_1^\dagger\Phi_2+{\rm h.c.}] \nonumber
\end{align}
\begin{align}
+& \frac{\lambda_1}{2}(\Phi_1^\dagger\Phi_1)^2
+\frac{\lambda_2}{2}(\Phi_2^\dagger\Phi_2)^2
+\lambda_3(\Phi_1^\dagger\Phi_1)(\Phi_2^\dagger\Phi_2)\nonumber\\
+&\lambda_4(\Phi_1^\dagger\Phi_2)(\Phi_2^\dagger\Phi_1) 
+\frac{1}{2}[\lambda_5~(\Phi_1^\dagger\Phi_2)^2 +~{\rm h.c.}]\nonumber\\
+&\big[ (\lambda_6(\Phi_1^\dagger\Phi_1)
+\lambda_7(\Phi_2^\dagger\Phi_2))
\Phi_1^\dagger\Phi_2+{\rm h.c.}\big] \,. \label{pot1}
\end{align}
Following the hermiticity of the scalar potential, \(m_{11}^2\), \(m_{22}^2\) and \(\lambda_{1,\ldots4}\) are real parameters
whereas \(m_{12}^2\), \(\lambda_{5,6,7}\) can be complex.
Assuming the CP-conserving version of the 2HDM, \(m_{12}^2\), \(\lambda_{5,6,7}\) and the VEVs of the fields \(\Phi_i\) are real parameters.
As a consequence of extending the discrete \(Z_2\) symmetry to the Yukawa sector in order to avoid Flavour Changing Neutral Currents (FCNCs) at tree level,
\(\lambda_{6,7}=0\), whereas the mass term \(m_{12}^2\) breaks the symmetry in a soft way.
The different transformations of the quark fields under the \(Z_2\) symmetry lead to four structures of Higgs-fermions interactions:
in Type-I only one doublet couples to all fermions;
in Type-II one of the doublets couples to the up quarks while the other doublets couples to the down quark;
in Type-X (or Lepton specific)  one of the doublets couples to all quarks and the other couples to all leptons;
in Type-Y (or Flipped) one of the doublet couples to up-type quarks and to leptons and the other couples to down-type quarks.

The Yukawa Lagrangian can be written in the form 
\begin{align}
-{\mathcal L}_Y=&+\sum_{f} \left[\left(\frac{m_f}{v}\xi_h^f \bar f fh+\frac{m_f}{v}\xi_H^f \bar f fH-i\frac{m_f}{v}\xi_A^f \bar f \gamma_5fA \right) \right]\nonumber\\
&+\frac{\sqrt 2}{v}\bar u \left (m_u V \xi_A^u P_L+ V m_d\xi_A^d P_R \right )d H^+ +\frac{\sqrt2m_ll\xi_A^l}{v}\bar\nu_L l_R H^+
+\text{h.c.},\label{Eq:Yukawa}
\end{align}
where $m_f$ is the relevant fermion mass, $P_{L,R}=(1\pm \gamma_5)/2$ and $V$ denotes the Cabibbo-Kobayashi-Maskawa (CKM) matrix.
Thus, in the absence of FCNCs, the 
Higgs-fermion couplings are flavour diagonal in the fermion mass basis and depend only on the mixing angle, $\beta$, in the alignment limit.
where the coefficients $\xi_{h_i}^f$ are  interpreted as the ratio of the Higgs boson couplings to the fermions with respect to the SM values,  which are defined in the alignment limit in Tab.~\ref{tab:1}, limitedly to the case of the $H$ and $A$ states, which are of interest here.
\begin{table}[t!]
\centering
 \begin{tabular}{| c | c |c| c | c |} 
 \hline
 Couplings & Type-I & Type-II & Type-X& Type-Y \\ [0.5ex] 
 \hline\hline
 $\xi^u_A $ & $\cot\beta$ & $\cot\beta$ & $\cot\beta$ & $\cot\beta$\\ [0.5ex] 
 $\xi^u_H $ &  -$\cot\beta$&  -$\cot\beta$& -$\cot\beta$ &-$\cot\beta$ \\[0.5ex] 
 \hline
 $\xi^d_A $&  -$\cot\beta$& $\tan(\beta)$ & -$\cot\beta$ & $\tan(\beta)$ \\[0.5ex] 
 $\xi^d_H$ &  -$\cot\beta$&  $\tan(\beta)$&  -$\cot\beta$& -$\tan(\beta)$\\[0.5ex] 
 \hline
 $\xi^l_A $& -$\cot\beta$ & $\tan(\beta)$ & $\tan(\beta)$ & -$\cot\beta$\\ [1ex] 
 $\xi^l_H$ & -$\cot\beta$ & $\tan(\beta)$ & $\tan(\beta)$ &-$\cot\beta$\\[0.5ex]
 \hline
 \end{tabular}
 \caption{Yukawa couplings of $A$ and $H$ in the alignment limit of the 2HDM types. } \label{tab:1}
\end{table}

Since we are interested in the two decays processes \AZH~and \HZA,
recall that the coupling of the heavy neutral Higgs scalar with the pseudoscalar and the gauge boson $Z$ in the 2HDM is given by:

\vspace{-0.75cm}
\begin{alignat}{2}
&g_{H AZ}: & \quad & \sin(\beta-\alpha).
\end{alignat}

\subsection{Theoretical and Experimental Constraints}
There are several theoretical and experimental constraints for the parameter points of the 2HDM to pass, {discussed} below.	
\begin{itemize}
	\item[\textbullet] Unitarity: various scattering processes  require that unitarity is conserved at the tree-level at high energy.
    The unitarity requirements in the 2HDM have been studied in~\cite{Kanemura:1993hm, Akeroyd:2000wc, Arhrib:2000is}.
	Sets of eigenvalues $e_i$ ($i-1, ... 12$) for the scattering  matrix of all Higgs and Goldstone bosons of the 2HDM are obtained as follows:
	\begin{eqnarray}
	&& e_{1,2} =  \lambda_3+2\lambda_4\pm 3 | \lambda_5| , \quad  \quad  e_{3,4} = \lambda_3\pm\lambda_4 , \quad e_{5,6} =  \lambda_3\pm|\lambda_5|,  \nonumber \\
	&&
	e_{7,8} = 3(\lambda_1+\lambda_2)\pm\sqrt{9(\lambda_1-\lambda_2)^2+4(2\lambda_3+\lambda_4|)^2},  \nonumber \\
	&&
	e_{9,10} = \lambda_1+\lambda_2\pm\sqrt{(\lambda_1-\lambda_2)^2+4|\lambda_5|^2},  \nonumber \\
	&&
	e_{11,12}  = \lambda_1+\lambda_2\pm\sqrt{(\lambda_1-\lambda_2)^2+4|\lambda_5|^2}.
	\end{eqnarray}
    We require all \(e_i\)'s to be less than 16\(\pi\) for each \(i=1,...12\).
	\vspace{-0.4cm}
\item[\textbullet]Perturbativity constraints \cite{Kanemura:1993hm,Branco:2011iw} implies that all that the quartic couplings of the scalar potential satisfy the condition \(|\lambda_i| \leqslant 8 \pi\) for each \(i=1,...5\).
	
	\vspace{-0.4cm}
	\item[\textbullet]Vacuum stability requires the scalar potential to be bounded from below~\cite{Gunion:2002zf} by satisfying the following inequalities:
	\begin{eqnarray}
	\lambda_{1,2}>0,  \,\,
	\lambda_3>- \sqrt{\lambda_1\lambda_2}, \,\,
	\lambda_3+\lambda_4-|\lambda_5|> - \sqrt{\lambda_1\lambda_2}.~~~
	\end{eqnarray}
	
	\vspace{-0.4cm}
	\item[\textbullet] EW Precision Observables (EWPOs) \cite{Haller:2018nnx}, such as the oblique parameters $S$ and $T$ \cite{Peskin:1991sw, Grimus:2008nb}, require a level of degeneracy between the charged Higgs boson state and one of the heavier neutral Higgs bosons. Here, we assume $m_{H^\pm} = m_{A}$  or $m_H$, as appropriate (see below), so that the $T$ parameter exactly vanishes in the alignment limit. 
	
	\vspace{-0.5cm}
\item[\textbullet] Exclusion limits at 95\% Confidence Level (CL) from Higgs searches at colliders (LEP, Tevatron and LHC) via HiggsBounds, version 5.3.2~\cite{Bechtle:2008jh, Bechtle:2011sb, Bechtle:2013wla} are enforced.
    Furthermore, the ATLAS Collaboration has set an upper limit at 95\% CL on the production cross section $\sigma$ of the $A$ state times its decay BR into $ZH\to l^+l^-b\bar b$, i.e., \(\sigma(A)\times {\rm BR}(A\rightarrow ZH \rightarrow l^+l^-b\bar b)\)~\cite{Aaboud:2018eoy}, that is not included in this tool, hence we have accounted for it separately.
	
	\vspace{-0.5cm}
\item[\textbullet] Constraints from the Higgs boson signal strength measurements are automatically satisfied as we assume $\sin(\beta-\alpha) =1$.	
	
	\vspace{-0.5cm}
\item[\textbullet] Constraints of {flavour physics observables,} namely, \(B \rightarrow X_s \gamma,~ B_{s,d} \rightarrow \mu^+\mu^-\) and \(\Delta m_{s,d}\)~\cite{Haller:2018nnx}.		
\end{itemize}

\section{Parameter scans and LHC sensitivity}

\subsection{The Scan}
A scan is performed over the parameter space of the 2HDM. In doing so, 
we use the program 2HDMC~\cite{Eriksson:2009ws}, firstly, to check the theoretical constraints as well as the EWPOs described above and, secondly,
to compute the Higgs BRs of each Higgs state, 
in particular those of  \AZH,~\HZA,~\Abb~and \Hbb. The
2HDMC code includes an interface to HiggsBounds, which is used to apply the aforementioned exclusion limits at 95\% CL from Higgs searches at LEP, Tevatron and LHC. Finally, 
the production cross sections of the heavy CP-even ($H$) and CP-odd ($A$) Higgs bosons,
at Next-to-Next-to-Leading Order (NNLO) in QCD, 
for both $gg\to H,A$ and $b\bar b\to A,H$, at the Centre-of-Mass (CM) energies of 13 TeV and 14 TeV,
are calculated using SusHi~\cite{Harlander:2012pb, Harlander:2016hcx, Harlander:2002wh, Harlander:2003ai}.

The first part of this study deals with the two production and decay processes \(pp \rightarrow H(A) \rightarrow ZA(H)\rightarrow b\overline{b}l^{-}l^{+}\).
The observed and expected confidence limits for all four types of Yukawa couplings in the 2HDM are produced at \(\sqrt{s}=13\)~TeV, with an integrated luminosity, $L$, of \(36.1~{\rm fb}^{-1}\),
by combining our calculations with the data from Ref.~\cite{Aaboud:2018eoy}. 
In the second part, we rescale the expected exclusion limit to the CM energy of \(\sqrt{s}=14\)~TeV,
with an integrated luminosity of \(300~{\rm fb}^{-1}\), by calculating the so called `upgrade factor' for both signals and backgrounds, while retaining the acceptance and selection efficiencies of the analysis at the lower $\sqrt s$ value. The change in energy will naturally affect signals and backgrounds differently. We treat the former by using SusHi (as intimated) and the latter by  using {MadGraph5, version 2.6.4}~\cite{Alwall:2011uj}. (For completeness, the
 background is considered to be any reducible or irreducible SM process that creates a pair of $b$-jets plus a pair of electrons or muons, as in Ref.~\cite{Aaboud:2018eoy}.)

\subsection{Numerical results}
In this study, we identify the lightest CP-even Higgs boson of the 2HDM as the observed Higgs state at the LHC, with $m_h=125$ GeV, 
and assume \(\sin (\beta-\alpha) = 1\).

We scan over the following parameter range:
\begin{equation}
\begin{gathered}
     m_{h} = 125~\text{GeV},~\sin(\beta-\alpha)=1, 0 < m_{12}^2 < 2\times 10^5~{\rm GeV},\\
     130 {\rm GeV} < m_X < 700~\text{GeV},~m_X \geq m_Y + 100~\text{GeV},\\
     m_X, m_Y~\text{chosen at}~10~\text{GeV intervals.}\\
     \tan(\beta)\in
         \begin{cases}
             \{1,~2,~3\}, & \text{if Lepton Specific}\\
             \{1,~5,~10,~20\}, & \text{otherwise}\\
         \end{cases}\\
\label{eq1}
\end{gathered}
\end{equation}
The set of values chosen for \(\tan(\beta)\), and the masses, align with the choices in~\cite{Aaboud:2018eoy}.
\begin{itemize}
    \item[\textbullet] For the process mediated by \AZH, we choose \(m_X~= m_A\), \(m_Y~= m_H\)  and \(m_{H^\pm} = m_A\). (Note that this choice is consistent with Ref.~\cite{Aaboud:2018eoy}.)
    \item[\textbullet] For the process mediated by \HZA, we choose \(m_X~= m_H\), \(m_Y~= m_A\)  and \(m_{H^\pm} = m_H\). (Note this choice is specular to that in Ref.~\cite{Aaboud:2018eoy}.) 
\end{itemize}

While an evident symmetry exists between the two cases, neither the constraints affecting the two processes nor their  sensitivity reaches should expected to be.
On the one hand, the role played by the heavy CP-even and CP-odd Higgs states of the 2HDM in both theoretical and experimental limits is different, owing to their different quantum numbers (and hence couplings).
On the other hand, their production and decay rates at the LHC are different despite leading to the same final states, including residual differences due to width effects entering their normalisation (but, as mentioned, not their kinematics), since, e.g., the $A$ state does not decay to $W^+W^-$ and $ZZ$ pairs while the $H$ state does and, conversely, the $A$ state decays to $Zh$ while the $H$ state does not.

\begin{figure}[t!]
	\centering
    \includegraphics[width=\textwidth]{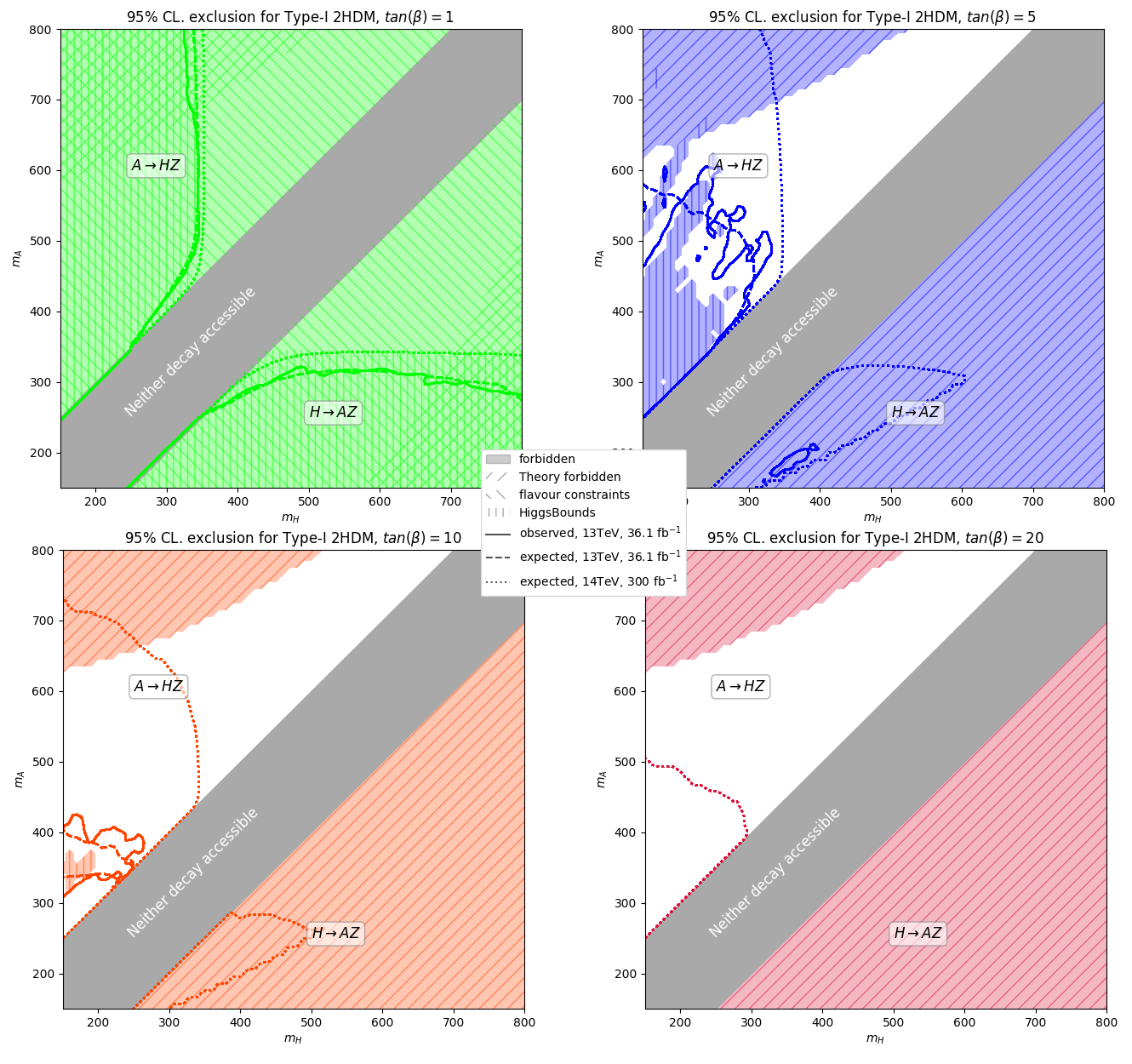}
    \caption{Exclusion limits at \(95\%\) CL in Type-I.
             The lines denoting expected and observed exclusion limits
             do not appear at all on some plots when the prediction never exceeds the 
             expected or observed  limit.}\label{fig1}
\end{figure}

\begin{figure}[t!]		
    \includegraphics[width=\textwidth]{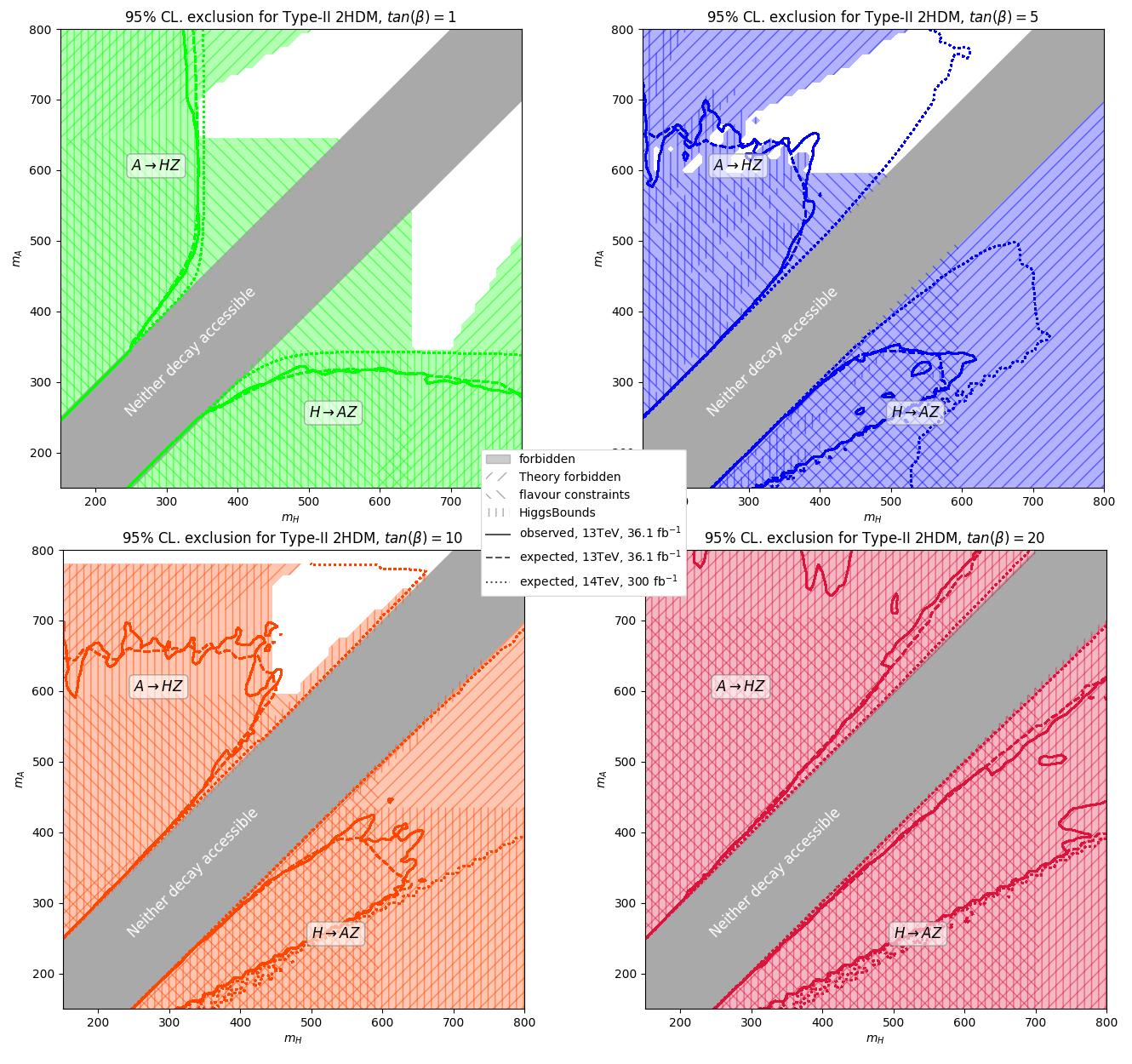}
    \caption{Like in Fig.~\ref{fig1} but for Type-II.}\label{fig2}
\end{figure}

\begin{figure*}[t!]	     
    \includegraphics[width=\textwidth]{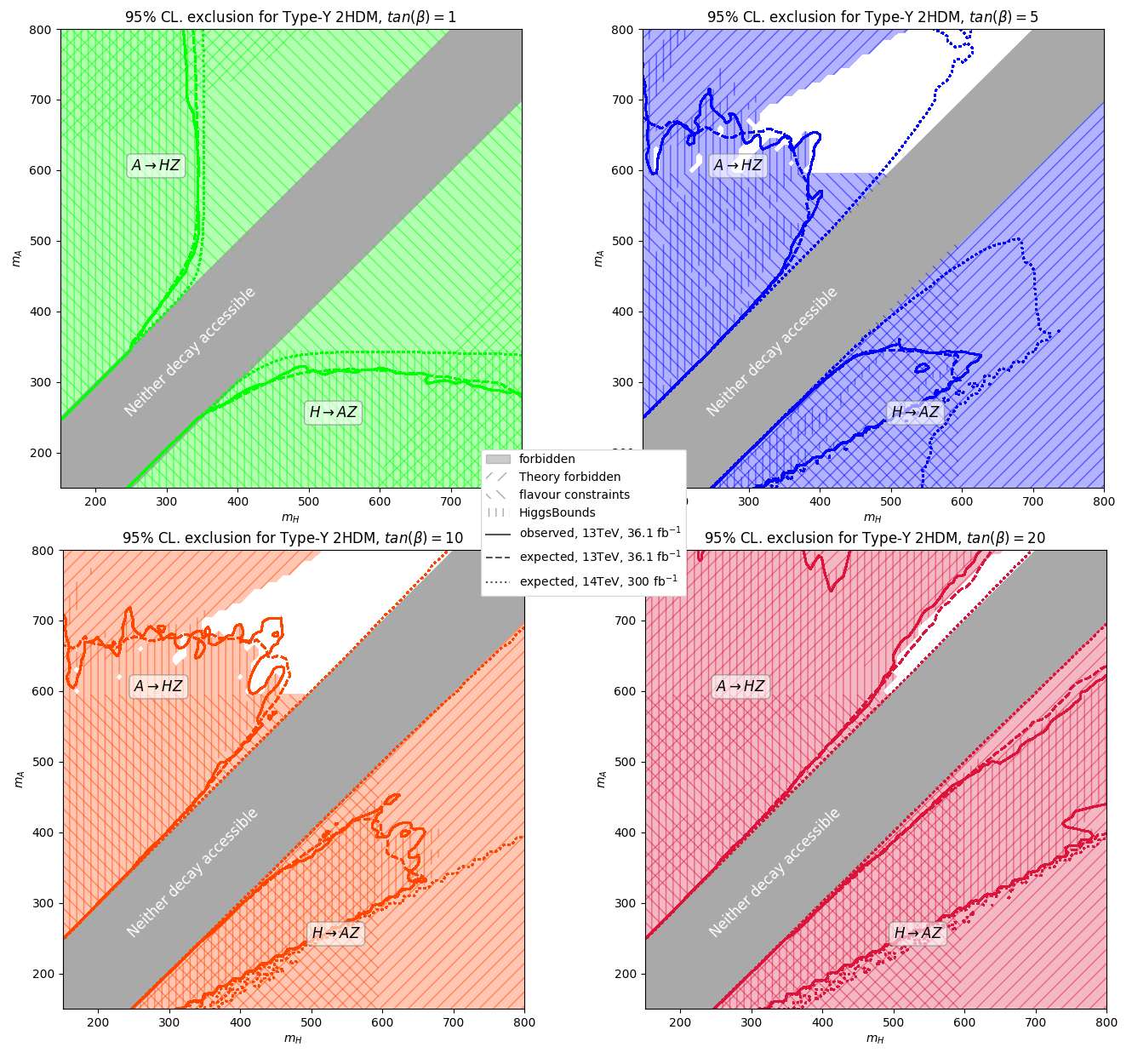}
    \caption{Like in Fig.~\ref{fig1} but for Type-Y (Flipped).}\label{fig3}
\end{figure*}

\begin{figure*}[t!]
	\centering
    \includegraphics[width=\textwidth]{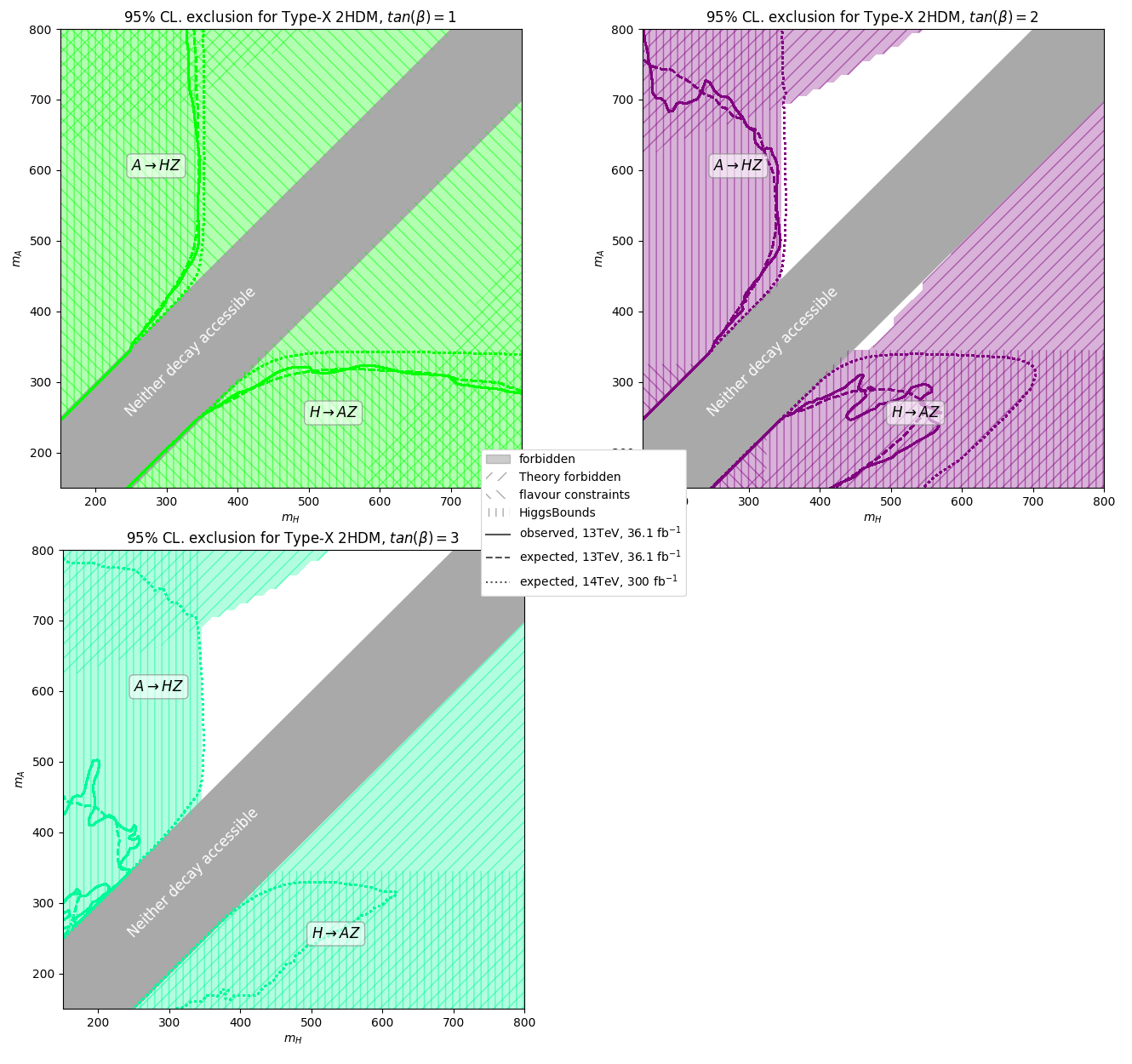}
    \caption{Like in Fig.~\ref{fig1} but for  Type-X (Lepton specific).}\label{fig4}
\end{figure*}

\begin{table}
\begin{tabularx}{\textwidth}{RCCCC}
      \toprule
      \textcolor{black}{\(\tan(\beta)\)}& \textcolor{black}{\(1\)} & \textcolor{black}{\(5\)} & \textcolor{black}{\(10\)} & \textcolor{black}{\(20\)}\\
      \toprule
      Type-I & \textcolor{red}{Flavour constraints} & \textcolor{blue}{Some masses} & \textcolor{blue}{Many masses} & \textcolor{red}{Low sensitivity} \\
      \hline
      Type-II &\textcolor{red}{Flavour constraints} & \textcolor{blue}{Some masses after upgrade} & \textcolor{blue}{Some masses after upgrade} & \textcolor{red}{Theory constraints}\\
      \hline
      Flipped &\textcolor{red}{Flavour constraints} & \textcolor{blue}{Some masses after upgrade} & \textcolor{blue}{Some masses after upgrade} & \textcolor{red}{Theory constraints}\\
      \toprule
      \textcolor{black}{\(\tan(\beta)\)}& \textcolor{black}{\(1\)} & \textcolor{black}{\(2\)} & \textcolor{black}{\(3\)} & \\
      \toprule
      Lepton specific & \textcolor{red}{Flavour constraints} & \textcolor{red}{Excluded by HiggsBounds} &  \textcolor{red}{Excluded by HiggsBounds} & \\
\end{tabularx}
\caption{Table summarising the findings in Figs.~\ref{fig1}~to~\ref{fig4}.
An overview of the possibility of each Yukawa type and value of \(\tan(\beta)\) is given.
Entries in red indicate that the combination has little or no mass combinations that are not forbidden while those in blue represent available parameter space accessible presently at Run 2  or after the upgrade of Run 3.}
\label{tab:summary}
\end{table}

After performing a scan over the parameter space delimited by Eq.~(\ref{eq1}), we compare the prediction of the model with the observed and expected limits given in Ref.~\cite{Aaboud:2018eoy}.
If the prediction exceeds the observed limit, then the parameter combination is excluded.
When the prediction exceeds the expected limit, we anticipate that the signal would be visible above background given the energies and luminosities available, hence,  {the experiment is sensitive to these parameters.} 

The choice of \(m^2_{12} = m_A^2 \tan(\beta) / (1 + \tan(\beta))^2\) enables us to reconstruct the exclusion limits at 95\%  CL given in Ref~\cite{Aaboud:2018eoy}.
However, this choice does not actually allow to satisfy theoretical constraints in all four types of 2HDM.
Therefore, we have dismissed it in our analysis.
In contrast, our choice of \(m_{12}^2\) above aims to simultaneously satisfy as many theoretical constraints as possible while affording one with significant parameter space amenable to experimental investigation.
Indeed, this is achieved by randomly sampling values of \(m^2_{12}\) between \(0\) and \(2\times 10^{5}\)~GeV for each point of the scan and selecting the one that passes most theoretical checks.

Figs.~\ref{fig1}~to~\ref{fig4} illustrate the outcome the scan for each
Yukawa type, \(\tan(\beta)\) and mass combination $(m_H,m_A)$.
Each figure provides results for one choice of Yukawa couplings
and each frame in each figure provides results at one value of \(\tan(\beta)\).
In the top left of each plot, where \(m_A > m_H+100\)~GeV, the decay \AZH{} is considered while 
in the bottom right of each plot, where \(m_H > m_A+100\)~GeV,   the decay \HZA{} is considered.
The corridor along the diagonal between these regions is coloured grey to indicate that neither decay is accessible.
If a combination of parameters is forbidden by theory, HiggsBounds or flavour constraints
then the corresponding area is filled with solid colour, conversely,
white areas pass all these checks and so are of interest. The hatching over the solid colour is used to indicate which of the checks
causes the corresponding parameter combination to fail.
There are three boundary lines drawn over the plots: 
these are the observed and expected \(95\%\) CLs for the ATLAS detector in its present state, \(13\) TeV and \(36.1~\text{fb}^{-1}\),
plus the expected 95\% CL for an upgraded LHC and ATLAS detector at \(14\) TeV and \(300~\text{fb}^{-1}\)\footnote{We neglect here to consider the case of $\sqrt s=13$ TeV and $L\approx140$ fb$^{-1}$, as it only improves marginally the present situation yet it would be make the plots far too crowded.}.
The model predictions exceed the 95\% CL inside the curve.

In Fig.~\ref{fig1} the parameter space with Type-I Yukawa couplings is shown.
The upper left plot shows that \(\tan(\beta) = 1\) is always forbidden by flavour constraints.
The upper right plot shows that there are many mass combinations that do not prevent the decay \AZH{} for \(\tan(\beta) = 5\),
but theory constraints forbid all mass combinations relevant to \HZA{}.
At \(\tan(\beta) = 5\) for \(13\) TeV (and \(36.1~\text{fb}^{-1}\)) the area of sensitivity (inside the expected curve) that is not excluded by observation (inside the observed curve) is very limited. 
At \(14\) TeV and \(300~\text{fb}^{-1}\), however, we expect many mass combinations to be testable that have not yet been excluded.
The lower left plot shows the behaviour at \(\tan(\beta) = 10\) to be similar to \(\tan(\beta) = 5\), i.e., 
everything is forbidden for \HZA{} by theory while for \AZH{} most combinations for which there is sensitivity have been excluded at \(13\) TeV
but \(14\) TeV offers even more possible parameter space than seen at \(\tan(\beta) = 5\).
Finally, in the lower right frame of Fig.~\ref{fig1}, the parameter space for \(\tan(\beta) = 20\) is shown.
The state of \HZA{} is unchanged, but now \AZH{} has no expected or observed exclusion at \(13\) TeV, i.e., these parameters are harder to probe.
With the upgrade to \(14\) TeV and 300 fb$^{-1}$ there is some sensitivity to \AZH{} at \(\tan(\beta) = 20\).

As might be expected, the behaviour of Type-II, shown in Fig.~\ref{fig2} and Type-Y, shown in Fig.~\ref{fig3}, is remarkably similar.
The upper left plot shows that \(\tan(\beta) = 1\) is forbidden by flavour constraints in all areas where there is sensitivity.
At \(13\) TeV  and 36.1 fb$^{-1}$ the upper right plot shows that the same can be said for \(\tan(\beta) = 5\), however,
after Run 3,  at \(14\) TeV and 300 fb$^{-1}$, there are many permitted mass combinations for \AZH{}. However, 
\HZA{} is excluded by theory.
The behaviour at \(\tan(\beta) = 10\), shown in the lower left plot, is much the same as for \(\tan(\beta) = 5\),
except more of the exclusion at \(13\) TeV and 36.1 fb$^{-1}$  is from observations provided by HiggsBounds.
Finally, in the lower right plot, \(\tan(\beta) = 20\) is shown to be excluded for almost all mass choices,
by multiple constraints.

In Fig.~\ref{fig4} the behaviour of the Type-X 2HDM is shown, at a set of \(\tan(\beta)\) values that differs from those previously considered.
For these Yukawa couplings and \(\tan(\beta)\) choices HiggsBounds excludes all areas inside the expected limits.
This remains true even after the end of Run 3.

Finally, Tab.~\ref{tab:summary} summarises our findings, highlighting that sensitivity only really exists for $5<\tan(\beta)<10$ and limitedly to the 2HDM Type-I, both at Run 2 and 3, and -II and -Y (or Flipped), but only at Run 3. The case of Type-X (or Lepton specific) is never accessible.

\section{Conclusions}
In summary, we have revisited an experimental analysis of the ATLAS Collaboration of the production and decay process $gg,b\bar b\to A\to ZH\to l^+l^-b\bar b$ performed at Run 2 with 36.1 fb$^{-1}$ of luminosity, which had been interpreted in terms of exclusion limits over the parameter space of the four types of the 2HDM, wherein the lightest Higgs state is identified with the SM-like Higgs boson discovered during Run 1 at the LHC with mass 125 GeV. Upon validating the ATLAS interpretation in our framework, though, we have discovered that their (fixed) choice of $m_{12}$, a mass parameter in the 2HDM Lagrangian that softly breaks an underlying $Z_2$ symmetry of the 2HDM to avoid FCNCs, yields parameter space configurations which are ruled out by theoretical requirements of model consistency. Hence, we have allowed this parameter to vary freely and subject the ensuing parameter space configurations to both the aforementioned theoretical constraints as well as those emerging from past and present experiments, thereby redrawing the actual sensitivity of such an experimental search to all four Yukawa types of the 2HDM, as a function of $\tan(\beta)$. In doing so, we have have also forecast the potential sensitivity of this channel to the 2HDM parameter space at the end of Run 3, assuming increased energy to 14 TeV and luminosity to 300 fb$^{-1}$.
This revealed some extended coverage of the 2HDM Type-I, -II and -Y (but not -X), 
especially for intermediate $\tan(\beta)$ values (say, between  5 and 10), with $m_A$ up to 800 GeV and $m_H$ up to 700 GeV.
This is somewhat beyond what is presently covered, i.e.,  up to 150 GeV or so in mass of either Higgs state, so as to justify further searches for this signature at the next stage of the LHC. Finally, we have recast the sensitivity of this analysis onto that of the channel  $gg,b\bar b\to H\to ZA\to l^+l^-b\bar b$. However, we have found that the complementary parameter space accessible this way (i.e., $m_H\ge  m_A+m_Z$) is actually entirely excluded already by existing theoretical and/or experimental constraints, so as to conclude that it is not warranted to pursue further this channel at the LHC, at least, not with a view to interpret it in the context of the standard four Yukawa types of the 2HDM\footnote{We finally note that analyses similar to Ref.~\cite{Aaboud:2018eoy}
performed by the CMS Collaboration exist \cite{Khachatryan:2016are,Sirunyan:2019wrn}. We have not used these for two reasons. On the one hand, they did not convey all the  information necessary to make  extrapolations to higher energies. On the other hand, they did not afford one with significantly different sensitivity to the 2HDM at present energies than what achieved by the ATLAS analysis ~\cite{Aaboud:2018eoy} that we have adopted as benchmark.}.

 
\section*{Acknowledgments}
RB and SS are supported by the Moroccan Ministry of Higher Education and Scientific Research MESRSFC and
CNRST: Project PPR/2015/6. HD-H is supported by the EPSRC CDT ``Next Generation Computational Modelling'' and acknowledges the use of the IRIDIS computing cluster at the University of Southampton. 
SM is supported in part through the NExT Institute and the STFC consolidated
Grant No. ST/L000296/1. 

\newpage

\bibliographystyle{plain}

\end{document}